# Shelling with MoS$_2$ – Functional CuS@MoS$_2$ Hybrids as Electrocatalysts for the Oxygen Reduction and Hydrogen Evolution Reactions

Avraham Bar-Hen[1,2], Ronen Bar Ziv*[,2], Tsion Ohaion-Raz[1,2], Amir Mizrahi[2], Simon Hettler[3], Raul Arenal,[3,4,5] and Maya Bar Sadan*[1,6]

[1] Department of Chemistry, Ben-Gurion University of the Negev, Beer-Sheva, 8410501, Israel

[2] Nuclear Research Center Negev, Beer-Sheva, 84190, Israel

[3] Laboratorio de Microscopías Avanzadas, Universidad de Zaragoza, 50018 Zaragoza, Spain

[4] ARAID Foundation, 50018 Zaragoza, Spain.

[5] Instituto de Nanociencia y Materiales de Aragon, CSIC-U. de Zaragoza, Calle Pedro Cerbuna 12, 50009 Zaragoza, Spain

[6] Ilse Katz Institute for Nanoscale Science and Technology, Ben-Gurion University of the Negev, Beer-Sheva, 8410501 Israel

* *barsadan@bgu.ac.il*

* *bronen@post.bgu.ac.il*

**Abstract**

The development of noble-metal free electrocatalysts is of high importance for clean energy conversion applications. $MoS_2$ has been considered as a promising low-cost catalyst for the hydrogen evolution reaction (HER), however its activity is limited by poor conductivity and low abundance of active sites. Moreover, its suitability as an effective catalyst for other reactions, in particluar the oxygen reduction reaction (ORR), was hardly explored to date. Herein, we show hybrid nanostructures of shelled CuS particles with $MoS_2$ layers, which produces several outcomes: The $MoS_2$ shell is strained and defective, and charge transfer from the core to $MoS_2$ occurs, enabling activation of the basal plane of $MoS_2$. Changing the feed ratio of the precursors led to control over morphology, such that the wrapping of the cores with the shell was continuously varied and characterized. We found an optimal hybrid structure, which provided high electrochemical active surface area and fast charge transfer kinetics, leading to improved activity not only towards HER (overpotential of 225 mV at 10 mA cm$^{-2}$), but also for the sluggish ORR (onset potential 0.87 V vs RHE).

## 1. Introduction

The development of new catalytic materials as an alternative to precious noble metals has attracted considerable attention for energy-related applications such as hydrogen generation from water splitting (hydrogen evolution reaction, HER) and the oxygen reduction reaction (ORR) in fuel cells and metal−air batteries.[1, 2] Yet many catalysts demonstrate high activity only towards one specific reaction, making the design of affordable bifunctional catalysts for both HER and ORR highly challenging. The ORR reaction occurs at the cathode of the alkaline fuel cell and has a large kinetic barrier due to a strong (di)oxygen double bond, which is the rate-determining step for the production of output energy from the initial $H_2$ fuel source; thus substantial energy input is required to initiate the reaction.[2, 3]

Low-cost molybdenum disulfide ($MoS_2$) can potentially replace precious metals as catalysts. However, the catalytic activity of pristine $MoS_2$ towards the HER is still lower than its theoretical potential,[4] and its suitability as an effective catalyst for other reactions, in particular ORR, has not been thoroughly investigated. The efficiency of $MoS_2$ is hindered by low conductivity and the low amount of active sites, e.g., edges and defects, with the basal planes being relatively inert.[5]

Strategies that can be used to overcome these limitations in $MoS_2$ and other transition metal dichalcogenides (TDMs) include increasing the density of their catalytic sites by creating defects,[6-9] modifying the morphology,[10, 11] or doping with other transition metals.[12-15] Another approach is to create hybrid materials by combining $MoS_2$ with other substances[11, 14, 16-25] such as core-shell materials, which are particularly effective for tuning the catalytic properties of layered nanomaterials. Thus far, several core materials have been selected for encapsulation by $MoS_2$.[26-30] The use of hybrid Au-$MoS_2$ core-shell structures as model photocatalysts[23, 24, 31] and electrocatalysts[18, 32] was recently reported. Our previous studies demonstrated that Au@$MoS_2$ nanoparticles enhance the catalytic activity by the charge transfer from Au to $MoS_2$ and by the strain within the $MoS_2$ lattice as it embeds the curved Au nanoparticle.[32] However, Au is a noble metal and is expensive, and practical usage requires structures that are affordable.

Other coinage-metal cores, such as Ag and Cu coated with $MoS_2$ layers, have been reported, and used for surface-enhanced Raman spectroscopy (SERS) applications.[33-36] More recently, a number of $MoS_2$-based heterostructures, in particular in core-shell morphology, such as $NiS@MoS_2$, $Mo_2C@MoS_2$, $CoS_2@MoS_2$, and $Ni_3S_2@MoS_2$ emerged as novel HER catalysts.[29, 36-40] Specifically, the $CuS@MoS_2$ system has gained much interest in the last year for various applications: hydrogen evolution,[37] photocatalysis,[41] and as anti-bacterial agents for wound healing.[42] These results demonstrate that coupling $MoS_2$ with other earth-abundant transition metal dichalcogenides is an effective strategy to enhance the HER performance of $MoS_2$. To date, the impact on the catalytic performance of the various parameters such as the nature of the core-shell interactions, the degree of wrapping of the core by the shell and the existence of free standing and extending $MoS_2$ structures, is still not well understood. Many of the structures include a geometry where the $MoS_2$ layers are orthogonal to the surface, and a relatively thick shell of $MoS_2$ is formed. This morphology correlates the enhancement of the catalytic activity with the 3D structure and porosity of the $MoS_2$ shell. It can be assumed that the core is more than a template to the $MoS_2$ shell,[32] and CuS core offers additional benefits due to its unique properties for electron capture and transport.[37] Our previous study showed that the structure and activity of $Cu_{2-x}S@MoS_2$ nanooctahedra, with a single $MoS_2$ shell, are closely connected with the atomic-scale structure of the interface.[19] Here, we aimed at determining on a fundamental level, what is the interaction of a few layers of $MoS_2$ with the CuS core, by changing the coverage and studying the effect on the catalytic activity of the structures. We report the preparation, characterization, and catalytic activity towards the HER and ORR of $CuS@MoS_2$ hybrid nanostructures. Varying the feed ratio of the precursors enabled us to control the hybrids produced and to eliminate undesired features, *i.e.* free standing $MoS_2$, partially wrapped and non-wrapped cores. The curved $MoS_2$ shell was found to be strained and defective, and charge transfer from core to $MoS_2$ shell occurred. These effects were found to activate the basal planes of $MoS_2$. The optimal hybrid structure showed high electrochemical surface area and fast charge-transfer kinetics that led to enhanced performance towards the HER, and additionally towards the ORR. Our well-defined hybrid structures pave the way for the design of efficient, durable, and cost-effective electrocatalysts for energy-related applications.

## 2. Materials and methods

### 2.1 Materials

Copper sulfate ($CuSO_4$, powder, 99.9% sigma aldrich), ammonium tetrathiomolybdate (($NH_4$)$_2MoS_4$ 99.97% sigma aldrich), molybdenum (VI) dioxide bis(acetylacetonate) (95% min STREM chemicals), dibenzyldisulfide (+98% Alfa Aesar), ethanolamine (99% Acros organics), toluene (99.7% Bio lab), Nafion D-521 (5% Alfa Aesar), Pt 20% on carbon black (Alfa Aesar), Sodium borohydride ($NaBH_4$, powder, 99% sigma aldrich). The water used for the synthesis was deionized water (DW) that was passed through a Milli-Q column by Millipore, with final resistance of 18.2 MΩ cm. All the chemicals were used as received without further purification.

### 2.2 Methods

Optical absorption measurements of the suspensions were obtained on a high-resolution Shimadzu 3600 Double-Beam UV-Vis-NIR spectrophotometer, using quartz cuvettes with 1 cm optical path. X-ray diffraction (XRD) was carried out using a Panalytical powder X-ray diffractometer (Empyrean) equipped with a position-sensitive X'Celerator detector using a θ-2θ scan with a Cu Kα radiation (λ=1.5405Å). XPS data were collected using X-ray photoelectron spectrometer ESCALAB 250 ultrahigh vacuum ($10^{-9}$ bar) apparatus with Al Kα X-ray source and a monochromator. The X-ray beam size was 500 μm. Survey spectra were recorded with pass energy of 150 eV and high energy resolution spectra were recorded with pass energy of 20 eV. To correct for charging effects, all spectra were calibrated relative to a carbon C 1s peak positioned at 284.8 eV. Processing of the XPS results was carried out using the AVANTGE program. For Raman measurements, a Horiba LabRam HR evolution micro-Raman system was used, equipped with a Synapse Open Electrode CCD detector air-cooled to -60°C. The excitation source was 532 nm laser with a power of 0.6 mW on the sample. The laser was focused with a x50 objective to a spot of about 1.3 mm. The measurements were taken with a 600 g mm$^{-1}$ grating and a 100 mm confocal microscope hole. Transmission electron microscopy (TEM) images were recorded using a JEOL 2100F microscope operated at 200kV. Aberration-corrected Scanning (S)TEM images were obtained with a Thermo Fisher Scientific Titan Low-Base microscope equipped with a high-brightness field emission gun operated at 300 kV using the high-angle annular dark field (HAADF) detector. Samples were deposited on an Al-grid, stored under vacuum and plasma-cleaned for 10s prior to insertion into the microscope to reduce contamination. Electron energy-loss spectroscopy (EELS) was performed using the Gatan Tridiem ESR 865 EEL spectrometer, working at 300 kV and additionally at 80 kV to improve the signal-to-noise ratio in the spectra. Analysis of the EELS data was performed with a custom Matlab program employing non-negative matrix factorization (NMF) to separate the S-contribution to the CuS in the core from the $MoS_2$ shell. Background-subtracted Mo-N (35 eV), Cu-M (74 eV), S-L (165 eV) and C-K (284 eV) edges were used for elemental mapping. Quantification of the Cu/S ratio was performed on only partially shelled cores without $MoS_2$ coverage on top and bottom side with respect to the electron-beam direction as checked by the absence of Mo EELS signal in the core region. Cu-M and S-L edges were used for quantification of the Cu/S ratio in the core nanoparticles by integrating the background-subtracted spectra intensity in a window of 30 eV and calculating the respective cross sections using different scripts provided in ref [43]. The Cu-L edge (932 eV) was used to ensure correct quantification using the unpronounced Cu-M edge.

## 2.3 Electrochemical measurements

A standard three electrode setup was used for all the electrochemical measurements. Graphite rod, Ag/AgCl and glassy carbon electrode coated with the examined catalyst were used as the counter, reference and working electrodes, respectively. Polarization curves and electrochemical impedance spectroscopy (EIS) measurements were recorded on an Ivium Technology Vertex Potentiostat/ Galvanostat (V74606) and analyzed using the Ivium Soft program. ORR performance was measured using rotating ring-disc electrode apparatus (RRDE-3A, ALS). Working electrodes were prepared as follows: ink was prepared by dispersing 2 mg catalyst and 1 mg Vulcan carbon black in a mixture of 415 μl deionized water, 104 μl ethanol and 31 μl Nafion solution (5%). The ink was homogenized by bath ultra-sonication for 2 min followed by probe sonicator (QSONICA 125W) at 40% amplitude in an interval mode of 15:5 seconds on:off cycle for 5 min. Then, 20 μl of the homogeneous ink was drop-casted onto a mirror-polished 3 mm glassy carbon electrode to form a final loading of ~1 mg·cm$^{-2}$. The electrode was dried at ambient temperature and was used as the working electrode for further electrochemical measurements. All the HER measurements were conducted in an Ar-saturated 0.5 M $H_2SO_4$ solution at ambient temperature. The electrolyte was bubbled with Ar gas for 15 minutes prior to the measurements to remove dissolved gases from the solution. During the electrochemical measurements, the head space of the cell was continuously purged with Ar gas. Each electrode was pre-treated with 12 CV cycles between -0.1 V and -0.65 V (vs Ag/AgCl) at a scan rate of 100 mV·s$^{-1}$. Polarization curves were recorded at a scan rate of 10 mV·s$^{-1}$ over the same potential range. EIS measurements were performed at the range of 120 to 220 mV overpotential from 10 KHz to 0.05 Hz (10 points per decade) with the amplitude of the sinusoidal perturbation fixed at 15 mV. All measurements were referred to the reversible hydrogen electrode (RHE) by using the relationship: E(RHE)= E(Ag/AgCl)+ E$^0$(Ag/AgCl)+ 0.059V × pH. The Tafel slope was calculated according to the Tafel equation: $\eta = a + b \log|j|$ Where b is the Tafel slope (V·dec$^{-1}$), η (V) denotes the applied overpotential, j(A·cm$^{-2}$) is related to the current density. The ECSA measurements were performed according to a procedure described elsewhere.[44] Cyclic voltammograms, scanning in 0.5 M H$_2$SO$_4$, from 0.1 V to -0.1 V (vs. RHE), were recorded at various scan rates (20, 40, 60, 80, 100, 120, 140, 160, 180 and 200 mV·s$^{-1}$) to estimate the double-layer capacitance (CdL). All the ORR measurements were conducted in an O$_2$-saturated 0.1 M KOH solution at ambient temperature. Each electrode was pre-treated with 12 CV cycles between 0.7 V and -0.6 V (vs Ag/AgCl) at a scan rate of 100 mV·s$^{-1}$. The LSV curves for ORR were obtained at room temperature and rotating speed of 1600 rpm at a scan rate of 25 mV·s$^{-1}$. The electron transfer number (n) was determined from the RDE plots at rotating speeds ranging 240-4200 rpm (scan rate: 10 mV·s$^{-1}$) using the Koutecky-Levich equation (Eq. 1).

$$(1) \quad \frac{1}{i} = \frac{1}{i_k} + \frac{1}{0.62 \times nFC(D)^{2/3}(v)^{-1/6}} \cdot \frac{1}{\sqrt{\omega}}$$

Where i and $i_k$ are the measured and kinetic current densities, respectively; $\omega$ is the rotation speed, n is the number of electrons transferred in ORR, F is the Faraday constant, C is the saturated concentration of O$_2$ in the solution, D is the diffusion coefficient of O$_2$, v is the kinetic viscosity of the electrolyte and k is the electron transfer rate constant. The constant 0.62 is used when the rotation speed is expressed in rad·s$^{-1}$.

### 2.4 Synthesis of the hybrid nanostructures

First, Cu nanoparticles were prepared by reducing Cu(II)SO$_4$ using NaBH$_4$. In a sealed flask purged with nitrogen, 75ml of aqueous solution containing CuSO$_4$ (1.3 mM) were stirred for 5 minutes to de-gas oxygen from the system. Excess of ice-cold NaBH$_4$ solution (5ml, 83 mM) was added at once to the flask and stirred for additional 30 min to produce brownish suspension of Cu nanoparticles. Under continuous stirring, 20 mL (NH$_4$)$_2$MoS$_4$ solution (0.25 to 2.50 mg·L$^{-1}$) was added slowly to the Cu nanoparticles suspension for the MoS$_4^{-2}$ to attach to their surface. The total volume in the flask was 100 ml, i.e. the final $(NH_4)_2MoS_4$ concentration was in the range of 0.05-0.50 mg·L$^{-1}$. The suspension was stirred overnight and then centrifuged (12000 RPM) for 40 minutes. The supernatant was discarded and then the solid was washed with a mixture of aqueous ethanol (1:2) several times to remove excess thiomolybdate. The wet powder of the as-obtained Cu@MoS$_4^{2-}$ particles was placed in a quartz ampoule and dried completely overnight at 60$^0$C in vacuum. Thereafter, the ampoule was purged with Ar and sealed under vacuum (~10$^{-4}$ torr). The sealed ampoule was placed in an oven, such that a temperature gradient of 88$^0$C was formed along the ampoule: the ampule's edge with the particles was heated to 516$^0$C and the opposite edge was kept at 428$^0$C. the temperature gradient allowed for accumulation of the decomposition products (H$_2$S, S, NH$_3$) at the cold end. After 30 min, the tube was slowly cooled down to room temperature and the product was collected and characterized.

### 2.5 Synthesis of free-standing MoS$_2$

The reaction was carried out in a three neck round bottom (RB) flask connected to a Schlenk line apparatus. Molybdenum dioxide bis(acetylacetonate) (0.4 mmol, 131.27 mg) and dibenzyl disulfide (0.8 mmol, 98.4 mg) were placed in the flask. Ethanolamine (30 ml) was added to the flask and the reaction mixture was degassed under vacuum at 110$^0$C for 30 min. Then the flask was filled with nitrogen. The temperature of the reaction mixture was increased to 300$^0$C and was maintained for the next 60 min. After the completion of the reaction, the reaction mixture was allowed to cool to room temperature. The product was collected by centrifugation. The product was washed twice with toluene and ethanol. The obtained product was stirred overnight in 5 ml toluene, 5 ml ethanol and 2 ml ethanolamine mixture to remove the ligands. After stirring, the sample was collected by centrifugation, washed with toluene and vacuum-dried for 5 hours.

## 3. Results and discussion

A set of samples with various initial thiomolybdate concentrations was prepared, varying from 0.05-0.50 mg·L$^{-1}$ thiomolybdate, denoted CuS@MoS$_2$-0.05 to CuS@MoS$_2$-0.50. **Figure 1** shows TEM images of the core-shell hybrids, and SEM images are available in the supporting information (**Figure S1**). Catalysts prepared with lower ammonium tetra-thiomolybdate, (NH$_4$)$_2$MoS$_4$, concentrations showed low coverage of MoS$_2$ shells around the core, such that some structures remained bare and others were only partially wrapped (**Figure 1a-b**). As the thiomolybdate ligand also serves as stabilizer during the synthesis, its low concentration led to aggregation of the Cu structures. In contrast, high thiomolybdate concentration (0.50 mg·L$^{-1}$) resulted in multilayered MoS$_2$ shell and loosely attached MoS$_2$ structures (**Figure 1d**). **Table S1** summarizes the effect of thiomolybdate concentration in the reaction mixture on the products homogeneity, where a concentration of 0.30 mg·L$^{-1}$ (**Figure 1c**) provided the conditions to obtain well-wrapped core-shell structures with negligible free standing MoS$_2$ (< 5%).

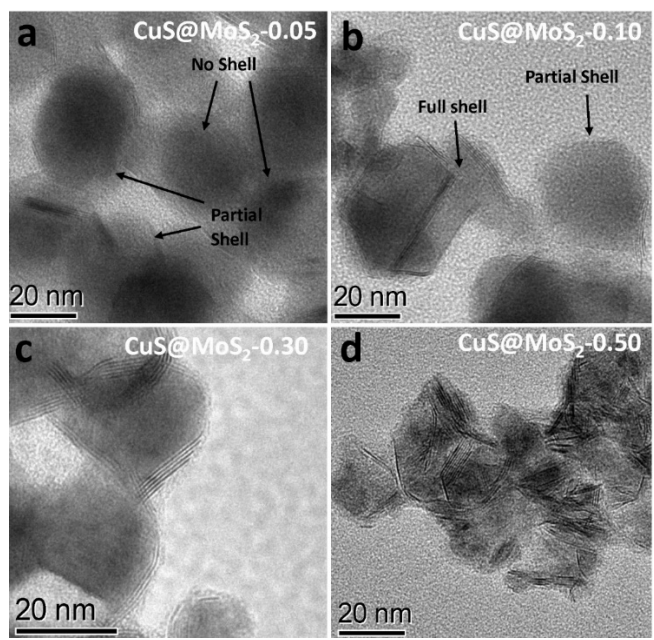

**Figure 1**. TEM images of CuS@MoS$_2$ with various thiomolybdate concentrations used in the synthesis (a) 0.05, (b) 0.10, (c) 0.30 and (d) 0.50 mg·L$^{-1}$.

Powder XRD patterns (**Figure 2a**) contain peaks of both $MoS_2$ and CuS (covellite),[45, 46] without evidence of metallic Cu. The Cu cores were sulfidized to CuS, potentially due to the release of sulfur during the decomposition of thiomolybdate and the formation of $MoS_2$.[47] According to this reaction path, the first stage is the adsorbance of $MoS_4^{2-}$ to the metallic Cu cores, forming the intermediate Cu@$MoS_4^{2-}$, which continues to react to form a $MoS_2$ shell by the reaction Cu@$MoS_4^{2-}$ + $2NH_4^+$ → $CuS_x$-$MoS_2$ + $NH_3$ + $H_2S$ + S. By comparing the peak area of the $MoS_2$ (002) and the CuS (103), we could analyze the ratio of the two phases (by mass). The increase in the content of $MoS_2$ within the hybrids is positively correlated with the concentration of the thiomolybdate in the feed solution (**Table S2**). It is also possible to calculate the strain in the $MoS_2$ lattice by the peak broadening (calculation of strain from XRD is explained in the Supporting Information, see **Table S3**).[29] Referring to the unveiling trend, rather than absolute numbers, the strain within the $MoS_2$ layer rises with additional $MoS_2$ coverage and reaches a maximum for the CuS@$MoS_2$-0.30. Raman measurements of the CuS@$MoS_2$ hybrids confirmed the formation of $MoS_2$ in all the samples (**Figure 2b**). The Raman measurements show that the separation between the $E^1_{2g}$ and $A_{1g}$ peaks increased from 23 $cm^{-1}$ to 25 $cm^{-1}$, which is generally correlated in the literature to additional layers of $MoS_2$, although precise comparison to the literature is limited due to different morphologies.[48] For CuS@$MoS_2$-0.30, a shift in the position of the $E^1_{2g}$ to below 380 $cm^{-1}$, is potentially indictive of S vacancies in the $MoS_2$ shell.[49-51] Moreover, redshift of the in-plane Raman $E^1_{2g}$ and $A_{1g}$ mode is sensitive to the built-in strain within the $MoS_2$ molecular layer.[16] Here, the red shift of the $E^1_{2g}$ and $A_{1g}$ modes was attributed to strain that originates from the curved $MoS_2$ shell,[16],[52] which is most pronounced for CuS@$MoS_2$-0.30. The overall picture is that additional $MoS_2$ layers provide fuller coverage of the cores. The $MoS_2$ layers are strained and potentially defected. At some point, excess $MoS_2$ forms extending materials that are more relaxed since they are not bound to the curvature of the cores.

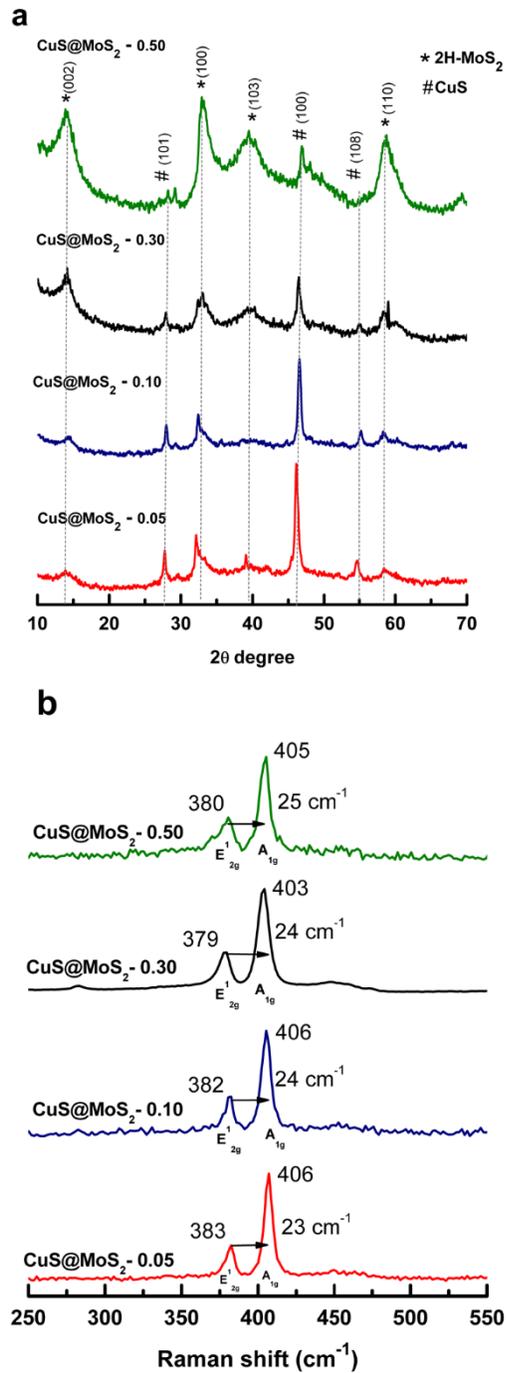

**Figure 2.** (a) Powder XRD pattern and (b) Raman spectroscopy of CuS@MoS$_2$ with various thiomolybdate concentrations.

Electrochemical measurements were performed to evaluate the catalytic activity of the various CuS@MoS$_2$ hybrids towards HER (**Figure 3a**) using linear sweep voltammetry (LSV) in a standard three-electrode setup under acidic conditions (0.5 M H$_2$SO$_4$). The results confirm that all the CuS@MoS$_2$ hybrids exhibit better performance with a significantly lower overpotentials (η) for a geometric current density of 10 mA·cm$^{-2}$ than Cu nanoparticles (431 mV) or free-standing MoS$_2$ (422 mV). It can be seen that among the hybrids sample set, the lowest overpotential was obtained by the CuS@MoS$_2$-0.30, which exhibited an overpotential of 225 mV at 10 mA·cm$^{-2}$. The HER kinetics were also estimated using the corresponding Tafel plots (**Figure 3b**), confirming a substantially lower Tafel slope (80 mV·dec$^{-1}$) of CuS@MoS$_2$-0.30 than the other CuS@MoS$_2$ hybrids or MoS$_2$ alone. Fast desorption and re-adsorption of H$_3$O$^+$ is particularly essential when designing catalysts for acidic conditions at high current densities, where large quantities of H$_2$ are generated. As shown in **Figure S3**, at high current densities, the superiority of CuS@MoS$_2$-0.30 as HER catalyst was maintained and becomes even more significant.

An effective electrocatalyst requires fast charge transfer kinetics, without which the desired electrochemical process will be hindered. To further study the electrode kinetics, electrochemical impedance spectroscopy (EIS) was applied on the various CuS@MoS$_2$ hybrids, where the interfacial charge transfer resistance (R$_{ct}$) can be represented by the semicircle fit in a Nyquist plot **(Figure 3c).** The CuS@MoS$_2$-0.30 exhibited the smallest electron transfer resistance **(Figure 3c)**, which leads to better catalytic activity in the HER process. The Tafel slope which is obtained by EIS from the linear fit of the plot of log (R$_{ct}^{-1}$) versus η (**Figure S4)** excludes the contribution of catalyst resistance, in contrast to that derived from the LSV curves.[12, 53] Therefore, similar values of the Tafel slope obtained by EIS and by the LSV indicate improved conductivity of the catalyst. Previous work has shown that pristine TMDs exhibit a difference between these two values, attributed to their electronic resistance [12, 53, 54]. In our study, the CuS@MoS$_2$-0.30 hybrid shows similar values of the Tafel slope values, indicating improved conductivity, unlike the other hybrids where partially shelled core or excess MoS$_2$ is present (**Figure S4**).

The electrochemically active surface area (ECSA), which is proportional to the electrochemical double-layer capacitance ($C_{dl}$) of an electrocatalyst, was measured by cyclic voltammetry at various scan rates[10, 15] (**Figure S5**). As shown in **Figure 3d**, the $C_{dl}$ of CuS@MoS$_2$-0.30 is fourfold larger than that of free standing MoS$_2$ and is the highest in the sample set of the hybrids, indicating the high fraction of exposed effective active sites, in according to the superior activity of the CuS@MoS$_2$-0.30 among the hybrid samples. Its stability after continuous CV cycles was also confirmed (**Figure S6**).

These results emphasis the strong dependency of the hybrids' composition and morphology on the synthesis conditions and the strong impact they have on the catalytic activity. We found that the cores should be sufficiently wrapped, to ensure the formation of a strained MoS$_2$ surface, which potentially contains also lattice defects and S vacancies. Excess MoS$_2$ results in relaxed, loosely connected MoS$_2$ material, which was less active as HER catalyst. Insufficient amount of MoS$_2$ shelling results in bare core surface and under-developed strain in the MoS$_2$ shell. The methodical and accurate set of experiments described here unmasks the relationship of the structure-catalytic activity, which was so far often missing. The best HER results, obtained with CuS@MoS$_2$-0.30 which proved as the optimal catalyst in the set, are remarkably close to those obtained for similar Au@MoS$_2$ structures,[32] but with a substantially more affordable core material.

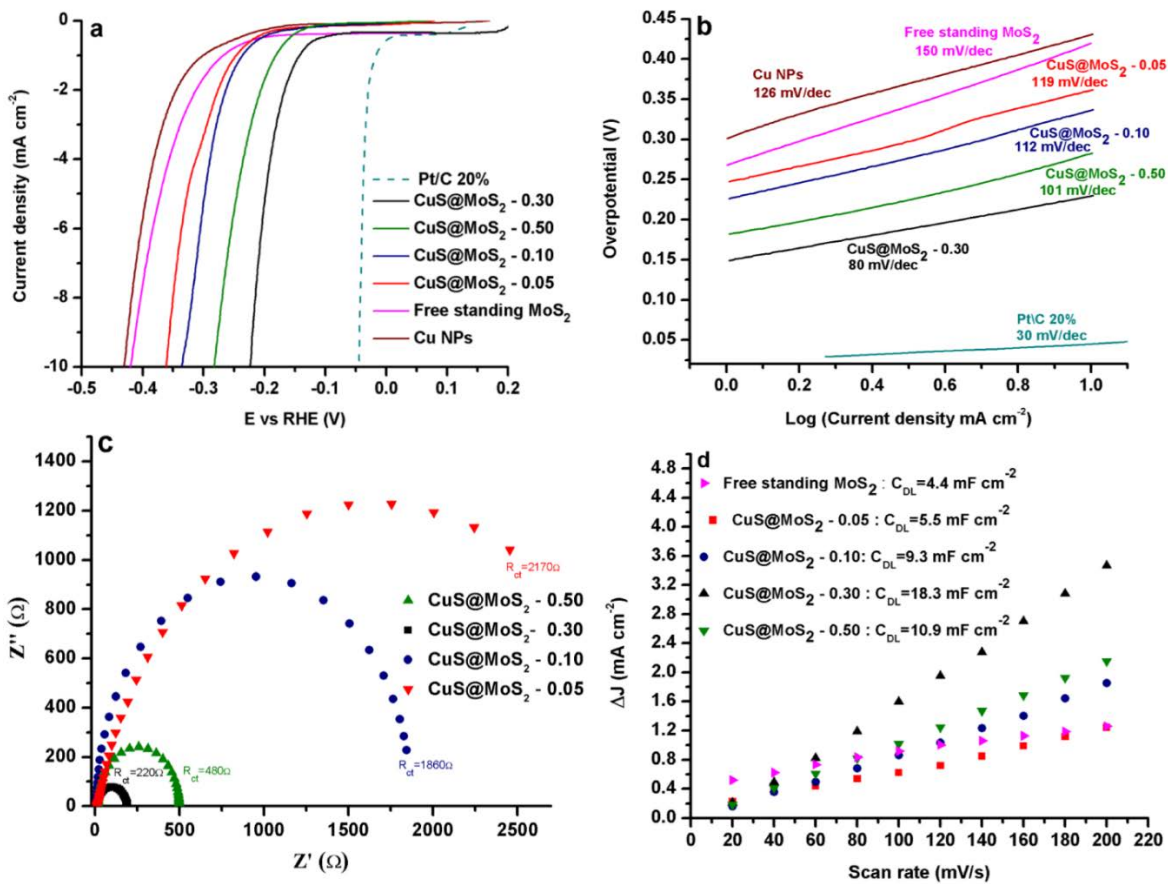

**Figure 3.** HER performance of the catalysts (a) LSV polarization curves of the various CuS@MoS$_2$ hybrids, free standing MoS$_2$, Cu nanoparticles and 20% Pt/C. (b) Corresponding Tafel slopes (c) The Nyquist plots (d) The electrochemically active surface areas (ECSA)

To gain more insights into the correlation between structure and catalytic activity, the CuS@MoS$_2$-0.30, which showed the best catalytic properties towards the HER, was further characterized. The high-resolution high-angle annular dark-field scanning TEM (HAADF-STEM) images shown in **Figure 4a-b** provide details of the shelling: The atomic layers of MoS$_2$ form a concentric shell with occasional defects such as dislocations and grain boundaries. The shelling is not uniform and, in some places, is not entirely complete. Statistical analysis of the TEM data provided an average core size of 20-30 nm and wrapping of 2-6 molecular layers. **Figure 4c** displays the lattice planes of a core nanoparticle as well as the atomic structure of the MoS$_2$ shell. The lattice plane distance (0.28 nm) of the core could correspond to the [103] direction of the covellite CuS phase. Electron energy-loss spectroscopy (EELS) analyses were performed to assess the elemental distribution within the CuS@MoS$_2$ hybrids. Spatial distribution of Cu (red), Mo (green), S (blue), and C (yellow) obtained from the EELS analysis of a hybrid nanoparticle is mapped in **Figure 4d**. The core-shell structure of the nanoparticles is clearly visible, with Mo only located in the shell region of the nanoparticle, and Cu confined to the core. The absence of Mo in the core region indicates an only partially shelled nanoparticle with no MoS$_2$ on the top and bottoms side with respect to the electron-beam direction. S is found both in the shell as part of the MoS$_2$, and in the core (**Figure S7**). Careful EELS quantification has been conducted, see all the details in the methods section, on several nanoparticles providing a Cu:S ratio of about 1:1, although the exact stoichiometry can differ by ~20%, suggesting a composition of Cu$_x$S with 0.8<x<1.2.

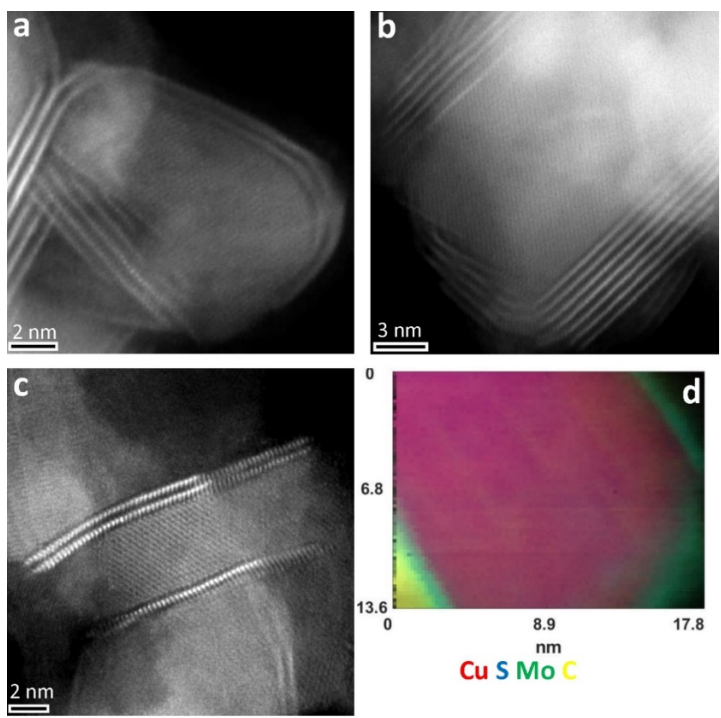

**Figure 4**. (a-c) High-resolution HAADF-STEM images of CuS@MoS$_2$-0.30 hybrids. (d) EELS analysis on an individual CuS@MoS$_2$-0.30 nanoparticle where the superposition of the elemental maps of Cu, S, Mo, and C (from the carbon support film on the Al TEM grid) is seen. The color code is shown below.

XPS measurements were carried out on CuS@MoS$_2$-0.30 to better understand the shelling process and to determine the chemical composition. The intermediate product (Cu@MoS$_4^{2-}$ or Cu-S$_x$-MoS$_{4-x}$), i.e. before thermal treatment, showed the existence of two different sulfur species in the XPS signal. One is attached directly to the Cu and is characterized by higher 2p binding energy leaving the unbounded sulfide unit in MoS$_4^{2-}$ with lower S$_{2p}$ binding energy (with a ratio of ~ 1:3 respectively, see **Figure S8** for detailed analysis). After thermal treatment, the final hybrid product showed the typical peaks of Mo 3d in MoS$_2$ (**Figure S9**). However, these peaks are shifted to lower binding energies by ~ 0.45 eV, and the Cu peaks also shifted towards higher energies, indicating charge transfer from the core to the MoS$_2$.

This observation supports the conclusion that electronic reconstructions occurred, which was recently correlated with improved electrocatalytic activity of core-shell hybrid structures.[32] Moreover, deconvolution of the Cu peaks unveils Cu$^0$/Cu$^{1+}$ species at 932.9 eV and 952.6 eV,[55] indicating that the conversion to CuS is not complete. The high-resolution spectrum of S 2p of the final product (**Figure S9**) includes two peaks, S 2p$_{3/2}$ at 162.6 eV and S 2p$_{1/2}$ at 163.8 eV, corresponding to S$^{2-}$ in MoS$_2$. In addition, a small peak was observed at higher binding energy (~165.2 eV) and was attributed to Cu-S at the interface between the MoS$_2$ layer and the core. The optical properties of an aqueous suspension of the annealed hybrid particles were measured by UV-Vis absorption. The spectrum contained peaks at 670 nm, 630 nm, and a broad peak at 400–550 nm, typical to the A, B, and C excitons of MoS$_2$, respectively (**Figure S10**).[56-59] The typical CuS absorption in the IR range shows a broad band around 1450 nm [60] and is present for CuS@MoS$_2$-0.30 (**Figure S10**) as well as for all the hybrid samples (**Figure S11**). While the XRD, HRTEM, EELS and IR absorption confirm that CuS is present at the core, it is more challenging to detect metallic Cu. Before annealing, the local surface plasmon resonance (LSPR) of metallic Cu nanoparticles was detected (**Figure S10**), but shelling with MoS$_2$ occasionally masks LSPR peaks, such as in the case of Au.[61] Selected area electron diffraction (SAED) of an assortment of CuS@MoS$_2$-0.30 particles unveiled metallic Cu reflections (**Figure S12**). Therefore, although most of the core is made of CuS, it is plausible that metallic Cu domains still exist in some of the particles. To summarize the findings so far, we showed that there is an optimal degree of $MoS_2$ shelling, that provides the adequate amount of lattice strain in $MoS_2$ and defective structural features, into which charge was transferred form the core. During the synthesis, the Cu cores are sulfidized to CuS, although some evidence show that metallic Cu still exists in the core, further from the surface. As a catalyst, this material offers better charge transfer kinetics, efficient catalytic sites and improved conductivity. The effect of shelling is therefore composed of a few contributing parameters, that should be delicately balanced to optimize the hybrids.[62]

Extensive research is currently being on design of nanomaterials as bifunctional non-noble-metal electrocatalysts, such as for both HER and ORR. Encouraged by the improved properties of the CuS@$MoS_2$-0.30 for HER, we further explored its activity as an electrocatalyst towards the ORR in alkaline conditions. It showed a substantial oxygen reduction peak for the experiment in $O_2$-saturated electrolytes, which was absent in Ar-saturated electrolytes (**Figure S13**). **Figure 5a** presents the LSV curves obtained at 1600 rpm in $O_2$-saturated 0.1 M KOH using a rotating disk electrode. For comparison, the LSV curves of free standing $MoS_2$ and commercial Pt/C were also recorded under the same conditions. CuS@$MoS_2$ exhibited excellent ORR activity, with ORR onset of 0.87 V versus RHE (half wave of 0.76 V), significantly more positive than free-standing $MoS_2$ (0.73 V, with relatively low current densities and half wave 0.64 V). While by our synthetic method we could not produce pure CuS particles as a reference sample, we measured the ORR activity of the sample with the lowest $MoS_2$ coverage, CuS@$MoS_2$-0.05, which contained 90% CuS by the XRD analysis (line 1, **Table S2**), and compared their activity with other published work on CuS (**Figure S14**).[63] The CuS@$MoS_2$-0.30 hybrid demonstrated higher performance than the reference samples, confirming the synergistic effect of the optimal core-shell construction. Moreover, the performance of CuS@$MoS_2$-0.30 exceeded other recently reported values for state-of-the-art MoS$_2$-based structures (for a broad comparison, see **Table S5**) and was comparable to the performance of the Pt/C electrode. Theoretical calculations (DFT) confirm our findings that MoS$_2$ is a poor ORR catalyst, but they also propose that Cu dopants in the MoS$_2$ monolayer provide optimal interaction with the ORR intermediate species (e.g., OOH, O, and OH).[64] Others have attributed the enhanced catalytic activity towards the ORR to the strained shell that offers optimized adsorption conditions for O$_2$, which is the first step of the ORR reaction and is crucial for its initialization.[29, 54]

The number of electrons transferred during the reduction ($n_e$) is also an important factor in evaluating the ORR activity of a catalyst. While four-electron reduction of O$_2$ to H$_2$O/OH$^-$ is the favorable process, two-electron reduction of O$_2$ to hydrogen peroxide (HO$_2^-$) often occurs, resulting in lower energy-conversion efficiency and degradation of the fuel cell. To estimate the electron transfer number ($n_e$), LSV data was obtained with RDE at different rotation speeds at a scan rate of 10 mV·S$^{-1}$, as shown in **Figure 5b** with the corresponding Koutecký-Levich (K-L) plot (at 0 V versus RHE) in **Figure 5c.** An electron transfer number of 3.95 was calculated using the K-L equation (see Supporting Information), which is very close to the commercially available Pt/C electrode (3.99).

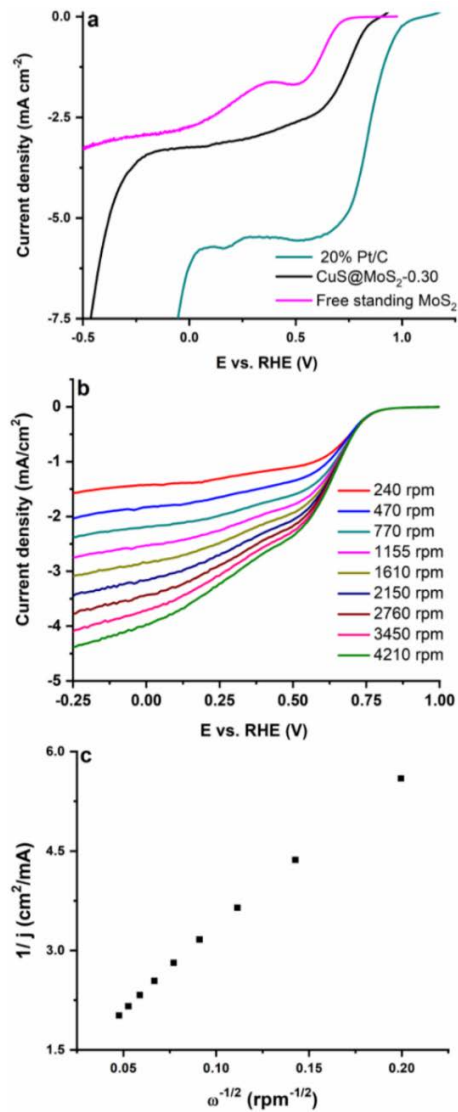

**Figure 5**: Electrocatalytic performance for the ORR in $O_2$-saturated 0.1 M KOH. (a) LSV measurements of the as-synthesized CuS@MoS$_2$-0.30 catalysts, free standing MoS$_2$, and commercial 20% Pt/C at 1600 rpm (scan rate: 25 mV·s$^{-1}$). (b) RDE experiment of CuS@MoS$_2$-0.30 at different rotation speeds from 240 to 4200 rpm at scan rate of 10 mV s$^{-1}$. (c) Koutecký-Levich (K-L) plot of CuS@MoS$_2$-0.30 (derived from b).

## Conclusions

In summary, we developed a versatile approach to synthesize CuS/MoS$_2$ core-shell hybrids and characterized them using various microscopy and spectroscopy techniques. An optimal CuS@MoS$_2$ hybrid was obtained by changing the concentration of thiomolybdate during the synthesis. This structure exhibited high electrochemical active surface area and fast charge transfer kinetics that facilitate enhanced electrocatalytic activity towards both HER (overpotential of 225 mV at 10 mA cm$^{-2}$) and ORR (onset potential 0.87 V vs RHE). The enhanced activity can be attributed to the activation of MoS$_2$ through optimizing the MoS$_2$ structural features as a shell, by forming basal plane strain, abundant defects and potential S vacancies, all commonly known to serve as active and effective catalytic sites; In addition, the charge transfer from the core to the MoS$_2$ shell that results in higher conductivity and lower charge transfer resistance between the electrocatalyst surface and reaction intermediates, contribute to the improved catalytic performance. As shown in this work, a non-noble conducting core is both affordable and effective means to promote the formation of active sites in the outer MoS$_2$ shell. By adjusting the ratio of the core and the shell materials, tuning and optimizing of the catalytic activity is further achieved. The obtained optimal catalysts are bifunctional both for HER and ORR. The insight gained here may serve as basis for further theoretical calculations on the relative importance of mechanical strains and electronic interactions that occur at the Cu-MoS$_2$ core-shell interface, and the implications of these effects on the catalytic properties of the MoS$_2$ shell.

## Declaration of Competing Interest

The authors declare that they have no known competing financial interests or personal relationships that could have appeared to influence the work reported in this paper.

**Acknowledgments**

This research was supported by the United States – Israel Binational Science Foundation (BSF), Jerusalem, Israel and the United States National Science Foundation (NSF) grant 2017642, and partly from the Israeli Atomic Energy Commission–Prof. A. Pazy joint foundation. The HRSTEM and EELS studies were conducted at the Laboratorio de Microscopias Avanzadas, Universidad de Zaragoza, Spain. S.H. acknowledges funding by German Research Foundation (HE 7675/1-1). R.A. gratefully acknowledges the support from the Spanish Ministry of Economy and Competitiveness (MINECO) and the MICINN through project grants MAT2016-79776-P (AEI/FEDER, UE) and PID2019-104739GB-I00/AEI/10.13039/501100011033, as well as from the European Union H2020 programs "Graphene Flagship" (881603) and "ESTEEM3" (823717).

**Supplementary Material.** UV-Vis, EELS and XPS data, and various electrochemical measurements.